\documentclass[fleqn,usenatbib]{mnras}

\usepackage{newtxtext,newtxmath}

\usepackage[T1]{fontenc}

\DeclareRobustCommand{\VAN}[3]{#2}
\let\VANthebibliography\thebibliography
\def\thebibliography{\DeclareRobustCommand{\VAN}[3]{##3}\VANthebibliography}

\newcommand{\soutdif}{\bgroup\markoverwith{\textcolor{magenta}{\rule[0.5ex]{2pt}{1pt}}}\ULon}

\newcommand{\soutPC}{\bgroup\markoverwith{\textcolor{cyan}{\rule[0.5ex]{2pt}{1pt}}}\ULon}

\usepackage{graphicx}	
\usepackage{amsmath}	
\usepackage{caption}
\usepackage{xcolor}
\usepackage{subcaption}
\usepackage[normalem]{ulem} 
\usepackage{orcidlink}

\title[Cosmic wallflowers as proto-globular clusters]{Too shy to spin? Cosmic wallflowers\thanks{\textbf{Wallflower} \textit{(noun)}: a person who has no one to dance with or who feels shy, awkward, or excluded at a party.}  as proto-globular clusters}
\author[F. van Donkelaar et al.] {Floor van Donkelaar$^{\orcidlink{0000-0002-7235-9747}}$,$^{1, 2}$\thanks{E-mail: fv294@cam.ac.uk}   
Lucio Mayer$^{\orcidlink{0000-0002-7078-2074}}$$^{3}$ and Pedro R. Capelo$^{\orcidlink{0000-0002-1786-963X}}$$^{3}$
\\
$^{1}$Institute of Astronomy, University of Cambridge, Madingley Road, Cambridge CB3 0HA, UK\\
$^{2}$Kavli Institute for Cosmology, Cambridge (KICC), University of Cambridge, Madingley Road, Cambridge CB3 0HA, UK\\
$^{3}$Department of Astrophysics, University of Zurich, Winterthurerstrasse 190, CH-8057 Z{\"u}rich, Switzerland}

\date{Accepted XXX. Received YYY; in original form ZZZ}

\pubyear{2026}

\begin{document}
\label{firstpage}
\pagerange{\pageref{firstpage}--\pageref{lastpage}}
\maketitle 

\begin{abstract}
We investigate the rotational properties of star-forming clusters at $z \sim 7.6$ in the high-resolution simulation \textsc{MassiveBlackPS}, focusing on two formation channels: clusters forming in galactic discs via gravitational instability and isolated circumgalactic systems, referred to as cosmic wallflowers, born out of cosmic filaments. Using stellar kinematics, we compare their rotational velocities, $v_{\rm rot}$, and rotational support, $v/\sigma$, to study whether formation environment leaves a clear dynamical imprint. We find that cosmic wallflowers preferentially show lower rotational velocities, whereas the identified disc clusters extend systematically towards higher $v_{\rm rot}$. In contrast, the two populations overlap substantially in $v/\sigma$, although the disc clusters show a modest preference towards larger rotational support. When combined with stellar surface densities, a subset of the low-$v_{\rm rot}$ cosmic wallflowers lie surprisingly close to the observed globular cluster population in the Milky Way, whereas disc clusters remain offset. Within the cosmic wallflower population, we identify two regimes: lower-density, weakly rotating systems that overlap with these globular cluster properties, and denser, more rotationally supported systems that likely follow a different evolutionary pathway, possibly linking them to the origin of massive black hole seeds at high redshift. We further find that the gas content correlates with this behaviour, with gas-rich cosmic wallflowers preferentially occupying this low-rotation regime. Together, these results suggest that formation environment is most clearly reflected in $v_{\rm rot}$, whereas gas content is linked to the variation within each population. In particular, weakly rotating, gas-rich cosmic wallflowers emerge as natural proto-globular cluster candidates, potentially evolving towards present-day systems through angular momentum loss and dynamical heating.
\end{abstract} 

\begin{keywords}
galaxies: high-redshift -- galaxies: clusters: general -- globular clusters: general -- methods: numerical  
\end{keywords}


\section{Introduction}\label{sec:introduction}

Stars predominantly form in all kinds of clustered environments, from short-lived open clusters to long-lived globular clusters (GCs), making their formation and evolution key to understanding galaxy assembly \citep[][]{Brodie:2006aa}. Recent observations with the \textit{James Webb Space Telescope} \citep[JWST;][]{Gardner:2006aa} have revealed compact, dense star clusters at $z \gtrsim 6$--10 \citep[e.g.][]{Vanzell:2023aa, Adamo:2024aa, Mowla:2024aa, bradac:2025aa}, with properties reminiscent of present-day GCs. These findings suggest that clustered star formation was a dominant mode of stellar assembly in the early Universe \citep[e.g.][]{adamo2025}.

At the same time, simulations have indicated that stellar clusters can form in diverse environments, including galactic discs \citep[e.g.][]{Kruijssen:2015aa, Pfeffer:2018aa, Mayer:2025aa} and more isolated, filamentary structures in the circumgalactic medium \citep[CGM; e.g.][]{Renaud:2017aa, vandonkelaar:2023aa, Donkelaar:2026aa}, referred to as ``cosmic wallflowers'' (CWs) in \citet{Donkelaar:2026aa}. In high-redshift galaxies, disc fragmentation is expected to proceed efficiently in gas-rich, turbulent systems \citep[e.g.][]{Dekel:2009aa, Tamburello:2015aa, Mayer:2016aa, Renaud:2021aa, vanDonkelaar:2022aa}. In addition, to help this disc fragmentation argument, recent observations with JWST and ALMA \citep[][]{Wootten_Thompson_2009} suggest that rotationally supported discs may already be common at these early epochs \citep[e.g.][]{Rizzo:2020aa, Rizzo:2021aa, Ferreira:2022aa, Ferreira:2023aa, Roman:2023aa, Danhaive:2025aa}. Alongside this, both observations and simulations indicate that compact stellar systems can also form outside galactic discs, for example in filamentary structures, galaxy interactions, or even dense outflows \citep[e.g.][]{Mandelker:2017aa, Mandelker:2018aa, Mowla:2022aa, Claeyssens:2023aa, Giunchi:2025aa, Whitaker:2025aa, Ong:2025aa}. These environments are expected to imprint different dynamical properties on forming clusters \citep[e.g.][]{Elmegreen:2010aa, Kruijssen:2012aa}, raising the question of whether some environments preferentially form clusters that are likely to survive as present-day GC analogues, and how their dynamical properties evolve over time.


\begin{table*}
\centering
\caption{Basic properties of the cluster samples used in this work. Values show the median and 16th--84th percentile range, measured at the final snapshot analysed here. We list the number of objects, $N$, the stellar and gas mass, $M_\star$ and $M_{\rm gas}$ (both computed within the stellar half-mass radius), the stellar half-mass radius, $r_{\rm half}$, the gas fraction, $f_{\rm gas}$ [defined as $f_{\rm gas} =M_{\rm gas}/(M_{\rm gas}+M_\star)$], the stellar rotational velocity, $v_{\rm rot}$, the stellar velocity dispersion, $\sigma$, and the stellar rotational support, $v/\sigma$. Definitions of these quantities are given in Section~\ref{ref:kin}. }
\label{tab:sample_properties}
{
\renewcommand{\arraystretch}{1.45}
\begin{tabular}{lccccccccc}
\hline
Sample 
& $N$ 
& $\langle \log_{10} M_\star \rangle$ 
& $\langle \log_{10} M_{\rm gas} \rangle$
& $\langle r_{\rm half} \rangle$ 
& $\langle f_{\rm gas} \rangle$ 
& $\langle v_{\rm rot} \rangle$ 
& $\langle \sigma \rangle$
& $\langle v/\sigma \rangle$ \\
& 
& $[{\rm M_\odot}]$ 
& $[{\rm M_\odot}]$
& $[{\rm pc}]$ 
& 
& $[{\rm km\,s^{-1}}]$ 
& $[{\rm km\,s^{-1}}]$
& \\
\hline
Cosmic wallflowers 
& 83 
& $6.31^{+0.98}_{-0.96}$ 
& $6.38^{+0.29}_{-0.39}$
& $2.52^{+1.38}_{-0.84}$ 
& $0.55^{+0.26}_{-0.35}$ 
& $14.9^{+37.2}_{-9.8}$ 
& $25.0^{+30.7}_{-14.5}$
& $0.85^{+0.54}_{-0.62}$ \\[0.7ex]
Disc clusters (Halo 0)
& {47} 
& {$6.83^{+0.46}_{-1.44}$}
& {$6.82^{+0.24}_{-0.66}$}
& {$2.61^{+0.30}_{-0.08}$} 
& {$0.49^{+0.36}_{-0.12}$} 
& {$34.2^{+46.8}_{-24.5}$} 
& {$53.7^{+19.6}_{-38.3}$}
& {$0.96^{+0.32}_{-0.51}$} \\
\hline
\end{tabular}
}
\end{table*}

Amongst the properties that should reflect, at least to some extent, the formation mechanism and original formation environment of GCs, is the kinematics. In particular, angular momentum, hence rotation of the clusters, can be related to properties of the flow such as shear, tidal  torquing, turbulence, or instabilities in their  parent gas flow. Provided this connection is at least partially maintained, kinematics can be used to provide constraints on the formation and long-term evolution of GCs. Observations of Galactic GCs have revealed that many systems exhibit internal rotation, suggesting that angular momentum is imprinted during formation and can persist over several dynamical times \citep[e.g.][]{Bianchini:2013aa, Bianchini:2018aa, Sollima:2019aa}. At the same time, theoretical studies show that clusters gradually lose angular momentum through two-body relaxation and tidal interactions, evolving towards more pressure-supported systems as they age \citep[e.g.][]{Tiongco:2017aa, Ernst:2007aa, Varri:2012aa}. Together, this implies that clusters forming with significant rotational support may evolve towards lower rotational velocity, $v_{\rm rot}$, and higher velocity dispersion, $\sigma$, over time, potentially approaching the kinematic regime of present-day GCs.

In this context, rotation may provide a key diagnostic of cluster survival. Numerical studies have shown that cluster disruption is strongly linked to both internal structure and dynamical state, with tidal shocks and interactions in dense galactic environments efficiently dissolving dynamically cold, disc-like systems \citep[e.g.][]{Gnedin:1997aa, Kruijssen:2015aa, Pfeffer:2018aa, Li:2017aa, Keller:2020aa}. Conversely, more pressure-supported clusters are expected to be more resilient against tidal stripping and evaporation \citep[e.g.][]{Fall:2001aa, Gnedin:2009aa, Kruijssen:2012aa}. Rotation may therefore be directly linked to cluster survivability, with systems forming in different environments following distinct evolutionary pathways set by their initial angular momentum and internal structure.

Nevertheless, while stellar kinematics can provide a direct probe of the dynamical state of clusters, it does not by itself capture the role of the baryonic component during formation. At early times, star-forming clusters are embedded in dense gas reservoirs \citep[e.g.][]{Lada:2003aa,Kruijssen:2012aa, Longmore:2014aa}, which can contribute significantly to the gravitational potential and mediate angular momentum transport \citep[e.g.][]{Grudic:2022aa}. The ability of clusters to retain or expel this gas is therefore expected to influence both their internal dynamics and subsequent evolution. In particular, models for GC formation suggest that gas retention in compact systems can play a key role in shaping their structure and the emergence of multiple stellar populations \citep[e.g.][]{Ercole:2008aa,Conroy:2011ab, bastian:2018aa,bobrick:2025aa}. This raises the extra question of whether the gas content of clusters is directly linked to their dynamical state, and whether it provides an additional pathway for identifying proto-GC candidates.

In this work, we investigate the rotational properties of star-forming clusters in different environments using the high-resolution isolated hydrodynamical simulation \textsc{MassiveBlackPS} \citep[][]{Mayer:2023aa}. We focus on clusters forming within galactic discs (Halo~0; \citealt{Mayer:2025aa}) and those forming more isolated in the CGM, i.e. the CWs \citep[][]{Donkelaar:2026aa}, and assess whether their internal kinematics differs systematically, and how these differences may relate to their subsequent dynamical evolution and survival.

\section{Methods}\label{sec:method}

We use the high-resolution isolated hydrodynamical simulation \textsc{MassiveBlackPS} \citep[][]{Mayer:2023aa}, run with the smoothed-particle hydrodynamics, $N$-body code \textsc{Gasoline2} \citep[][]{Wadsley:2017aa}. The simulation is based on \textsc{MassiveBlackHR} \citep[][]{Mayer:2023aa}, a cosmological zoom-in re-simulation of a highly overdense region selected from \textsc{MassiveBlack} \citep[][]{DiMatteo_et_al_2012,Feng_et_al_2014}, corresponding to a massive halo at $z \gtrsim 6$, reaching gas and dark matter particle masses (softenings) of $1.9 \times 10^4$ (142~pc) and $9.4 \times 10^4$~M$_{\sun}$ (241~pc), respectively. The isolated simulated region corresponds to the virial volume ($r_{\rm vir} \approx 39$~kpc) of the primary halo of \textsc{MassiveBlackHR} at $z \simeq 8$ and follows $\sim$60~Myr around a major ($\sim$1:1.2) merger, after performing particle splitting \citep[][]{Roskar_et_al_2015}. We focus on the final $\sim$6~Myr at $z \sim 7.6$, when the gas particle mass and softening are $2.4 \times 10^3$~M$_{\sun}$ and 2~pc, respectively, allowing us to resolve gas collapse and cluster formation on parsec scales. Star formation and feedback follow the recipes of \citet{Stinson:2006aa} with a \citet{Kroupa:2001aa} initial mass function. Cooling is computed for H and He in non-equilibrium with a redshift-dependent UV background \citep[][]{Haardt:2012aa}, while metal cooling assumes photo-ionization equilibrium with same background \citep[][]{Shen:2010aa,Shen:2013aa}, assuming no self-shielding \citep[][]{Pontzen:2008aa} in both cases \citep[see the discussion in][]{Capelo:2018aa}.

We analyse the rotational properties of two cluster populations identified in this simulation: (i) clusters forming via disc fragmentation in the central galaxy (Halo~0; \citealt{Mayer:2025aa}), and (ii) isolated circumgalactic CWs forming along filaments within the virial radius \citep[][]{Donkelaar:2026aa}. The basic properties of the two samples are summarised in Table~\ref{tab:sample_properties}.

\subsection{Kinematic measurements}\label{ref:kin}

\begin{figure}
    \centering
    \includegraphics[ trim={0cm 0cm 0cm 0cm}, clip, width=0.485\textwidth, keepaspectratio]{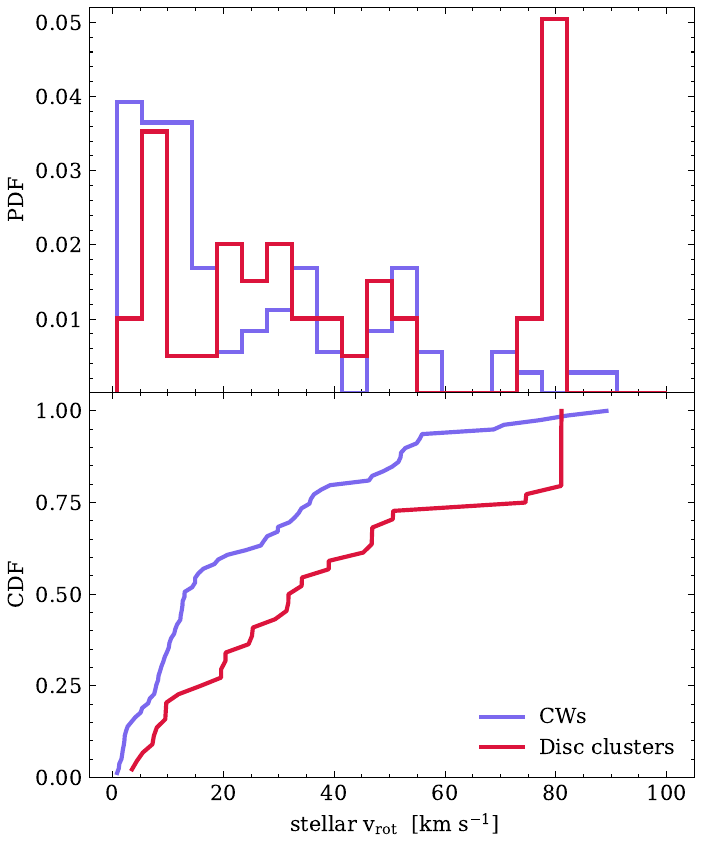}
    \caption{Stellar rotational velocity, $v_{\rm rot}$, distributions for the CWs (blue) and disc clusters (red) at the final snapshot ($z \sim 7.6$), shown as normalized PDFs (top panel) and CDFs (bottom panel).}
    \label{fig:Rot1}
\end{figure}

\begin{figure}
    \centering
    \includegraphics[ trim={0cm 0cm 0cm 0cm}, clip, width=0.485\textwidth, keepaspectratio]{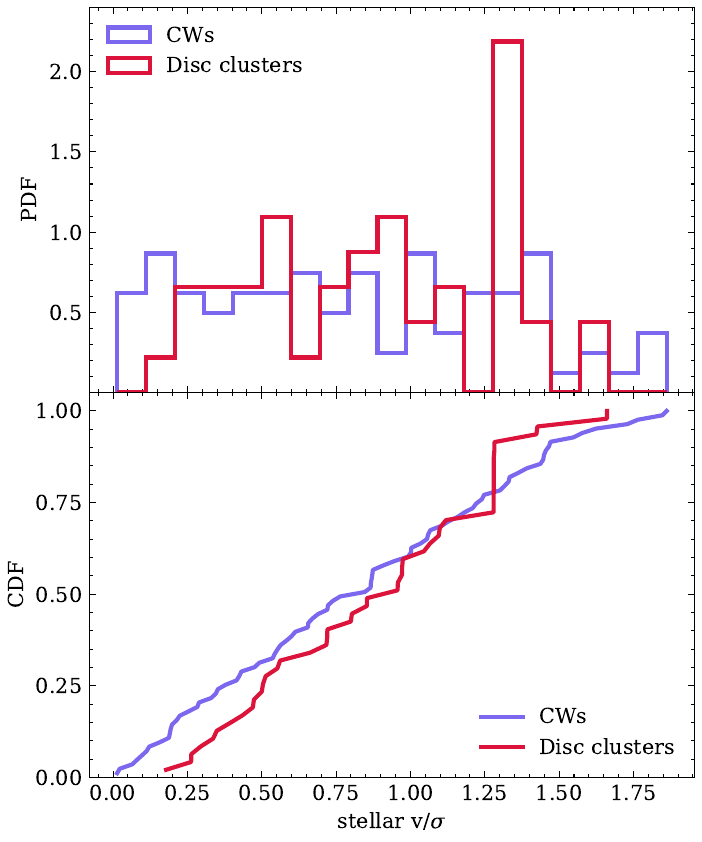}
    \caption{Stellar $v/\sigma$ distributions for CWs (blue) and disc clusters (red) at the final snapshot ($z \sim 7.6$), shown as normalized PDFs (top panel) and CDFs (bottom panel)}
    \label{fig:Rot2}
\end{figure}

For each cluster, we measure the internal kinematics within a compact spherical aperture based on the stellar half-mass radius, $r_{\rm half}$. For the disc-cluster sample, we re-centre each candidate on the local stellar-density maximum associated with the compact stellar structure and recompute $r_{\rm half}$ from the re-centred stellar distribution. The kinematics are measured within $2r_{\rm half}$, capped at the outer radius of the identified compact structure to minimise contamination from surrounding disc stars. The final re-centred sample contains 47 disc clusters, whereas the CW sample has 83 systems. The rotation axis is defined by the total stellar angular momentum,

\begin{equation}
\mathbf{J}_\star = \sum_i m_i , \mathbf{r}_i \times \mathbf{v}_i\,,
\end{equation}

\noindent where stellar particle positions are measured relative to the cluster centre identified by the halo finder, and velocities relative to the stellar centre-of-mass velocity. The stellar rotational velocity is then computed as the mass-weighted mean azimuthal velocity around this axis,

\begin{equation}
v_{\rm rot} = \frac{\sum_i m_i v_{\phi,i}}{\sum_i m_i}\,,
\end{equation}

\noindent whereas the velocity dispersion is defined as

\begin{equation}
\sigma =
\sqrt{
\frac{
\sum_i m_i (v_{\phi,i}-v_{\rm rot})^2
}{
\sum_i m_i
}
}.
\end{equation}

\noindent The ratio $v_{\rm rot}/\sigma$ therefore measures the internal rotational support of the stellar component and is not sensitive to the orbital motion of the cluster within the host halo. For brevity, we hereafter refer to the stellar $v_{\rm rot}/\sigma$ simply as $v/\sigma$.

For the gas component, we repeat the same measurement using the gas particles within the same aperture. We also compute the co-rotating mass fraction, $f_{\rm corot}$, defined as the fraction of mass with positive azimuthal velocity relative to the total angular momentum axis of the corresponding component,

\begin{equation}
f_{\rm corot} = \frac{\sum_{v_{\phi,i}>0} m_i} {\sum_i m_i}\,.
\end{equation}

\noindent Values of $f_{\rm corot}\approx 1$ indicate that most of the mass rotates in the same direction, whereas $f_{\rm corot} \approx 0.5$ corresponds to little net rotational coherence.

\section{Results}\label{sec:results}

Figures~\ref{fig:Rot1} and \ref{fig:Rot2} compare the stellar kinematics of the CWs and disc clusters at the final snapshot of the simulation ($z \sim 7.6$). The clearest difference is in $v_{\rm rot}$: the CWs preferentially show lower rotational velocities, whereas the disc clusters extend towards higher values. The cumulative distributions show the same systematic shift.

Figure~\ref{fig:Rot2} shows the same comparison as in Figure~\ref{fig:Rot1}, but now in terms of $v/\sigma$. In contrast to $v_{\rm rot}$, the two populations show substantial overlap in $v/\sigma$, although the disc clusters are slightly shifted towards higher values. The CWs span a wide range in $v/\sigma$, from dispersion-supported to rotation-dominated systems. A small subset of CWs reaches relatively high $v/\sigma$ values, which may at first appear surprising. Nevertheless, we have made sure that these systems are not associated with dark matter-dominated subhaloes, but instead remain strongly baryon-dominated with a baryon fraction\footnote{Defined as $f_{\rm bar} = (M_\star + M_{\rm gas})/(M_\star + M_{\rm gas} + M_{\rm DM})$.} $f_{\rm bar} \sim 1$, and therefore do not correspond to embedded mini-galaxies. A more detailed exploration of these clusters is shown in Appendix~\ref{sec:CF}. The narrow features in the disc-cluster PDFs around $v_{\rm rot}\sim80\,{\rm km\,s^{-1}}$ and $v/\sigma\sim1.3$ reflect several systems with similar kinematics falling within the same histogram bins. Their prominence is enhanced by the smaller disc-cluster sample, and we therefore do not interpret these peaks as evidence for a separate population.

\begin{figure}
    \centering
    \includegraphics[ trim={0cm 0cm 0cm 0cm}, clip, width=0.485\textwidth, keepaspectratio]{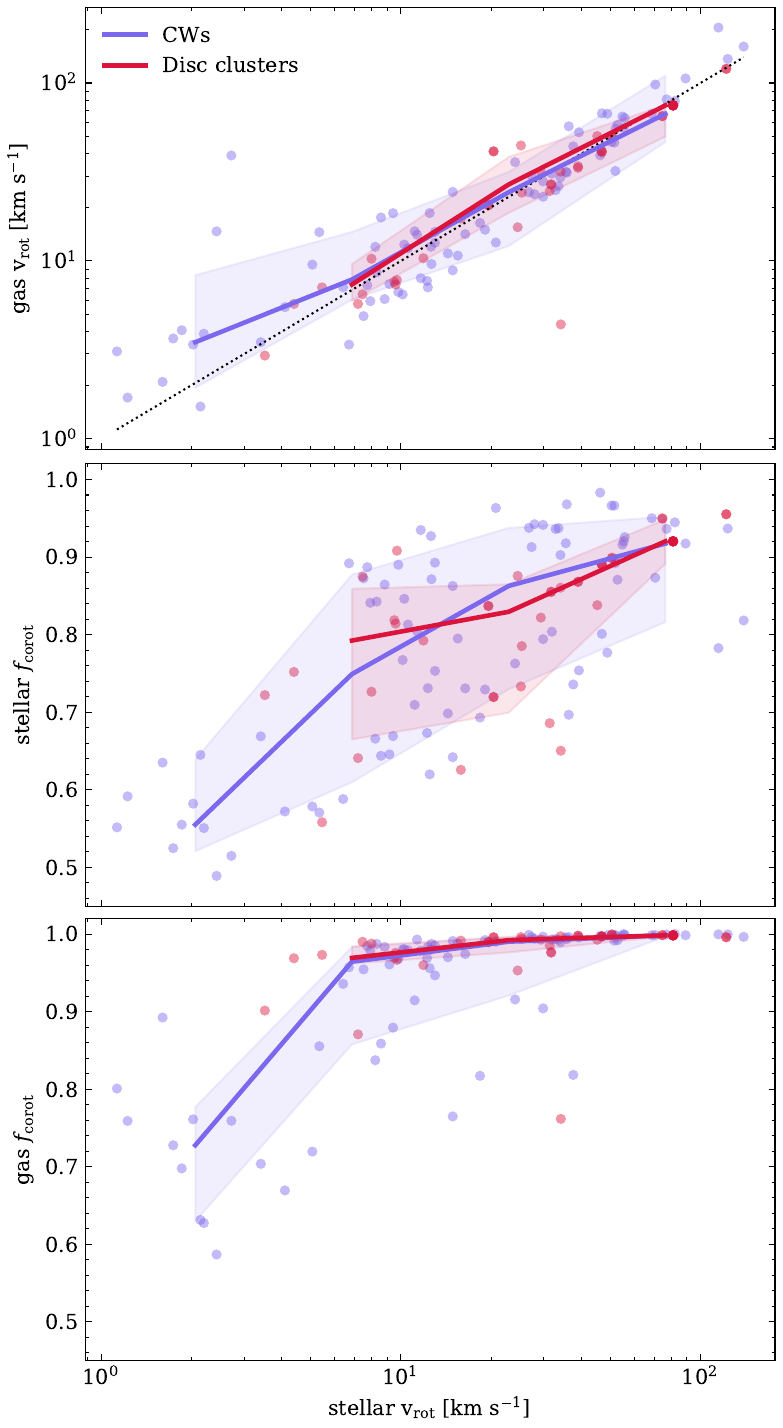}
    \caption{Rotational coherence as a function of stellar rotational velocity for CWs and disc clusters. From top to bottom, the panels show the gas rotational velocity with the dotted line marking the one-to-one relation, the stellar co-rotating mass fraction, and the gas co-rotating mass fraction. Solid lines show median trends in bins of stellar $v_{\rm rot}$, with shaded regions indicating the 16th--84th percentile range, and only shown for bins containing at least five objects.}
    \label{fig:vort}
\end{figure}

This diversity amongst the CWs is consistent with the presence of two physically distinct populations identified in \citet{Donkelaar:2026aa}, wherein the densest clusters are likely to undergo runaway stellar collisions and form intermediate-mass black hole (IMBH) seeds, whereas lower-density systems are expected to survive as bound star clusters. Although this separation is primarily driven by their internal density structure, the spread in $v/\sigma$ suggests that this physical distinction may also be imprinted in their dynamical state. This makes the low-$v_{\rm rot}$ CWs particularly interesting, as they may represent a subset of objects whose dynamical state is more closely aligned with the early stages of proto-GC formation.

\begin{figure*}
    \centering
    \includegraphics[ trim={0cm 0cm 0cm 0cm}, clip, width=0.999\textwidth, keepaspectratio]{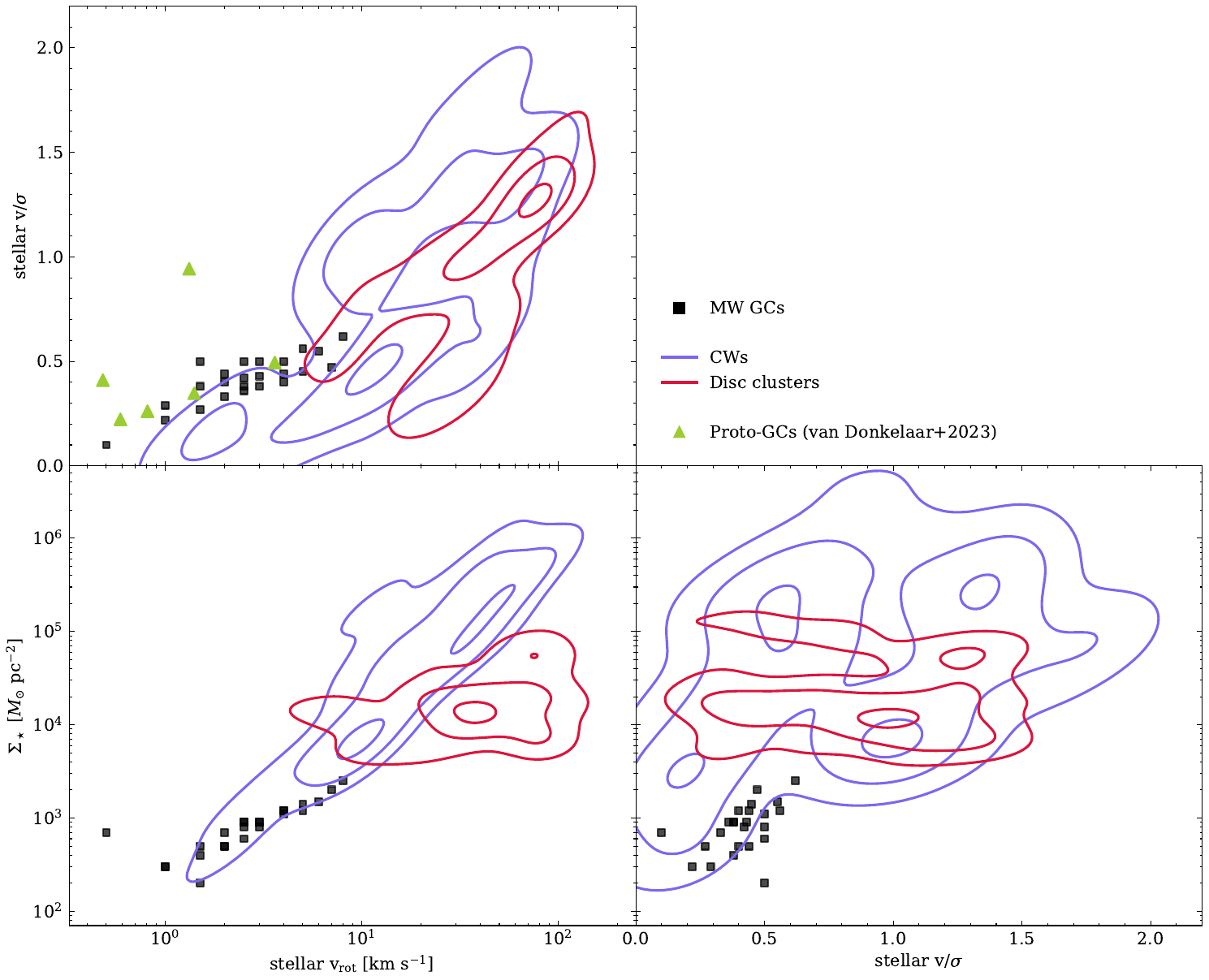}
    \caption{Stellar kinematic and structural properties of CWs (blue) and disc clusters (red) at $z \sim 7.6$. Black crosses show local GCs from \citet{Sollima:2019aa, Vasiliev:2021aa}, where $\sigma$ is measured at the stellar half-mass radius. The green triangles show the proto-GCs discussed in \citet{vandonkelaar:2023aa}.
    The top panel shows the stellar $v/\sigma$ as a function of stellar rotational velocity, $v_{\rm rot}$. The bottom left-hand panel shows stellar surface density as a function of $v_{\rm rot}$. The bottom right-hand panel shows the stellar surface density as a function of $v/\sigma$. Contours indicate the underlying distributions.}
    \label{fig:Rot3}
\end{figure*}

\subsection{Rotational coherence}

To further test whether the measured rotational velocities reflect coherent internal motion, we compare the stellar $v_{\rm rot}$ to several gas and stellar rotational-coherence diagnostics in Figure~\ref{fig:vort}. The CW population shows a tight correspondence between gas and stellar rotation, with a median gas-to-stellar rotational velocity ratio close to unity. Across the full CW sample, systems with higher stellar $v_{\rm rot}$ also tend to have higher gas $v_{\rm rot}$, indicating that the two components share a common angular-momentum structure. The CWs also have high co-rotating mass fractions, with median values of $f_{\rm corot}=0.81$ for the stars and $f_{\rm corot}=0.98$ for the gas. This indicates that the measured rotation of CWs is not driven by numerical noise or by a small number of pathological systems, but traces coherent gas-stellar rotational structure.

The disc clusters are shifted towards higher stellar rotational velocities and additionally show a clear correspondence between gas and stellar rotation. Their gas rotational velocities broadly follow the stellar component, close to the one-to-one relation, although with somewhat more scatter than in the CWs. The stellar and gas co-rotating fractions are also generally high, indicating coherent internal rotation in both components. The disc clusters also show somewhat larger scatter in their gas--stellar rotational properties than the CWs. The CWs have a median gas--star angle of only $\sim$$3^\circ$. The larger scatter likely reflects the more complex dynamical environment of the galactic disc, although both the stellar and gas components still show coherent rotation.

Again, this makes the low-rotation CWs particularly relevant for the comparison to GCs. These systems have $v_{\rm rot}<10\,{\rm km\,s^{-1}}$, placing them in the same low-rotation regime occupied by many present-day halo GCs. They also retain moderate stellar co-rotation and coherent gas rotation, with median co-rotating fractions of $f_{\rm corot}=0.64$ for the stars and $f_{\rm corot}=0.85$ for the gas, meaning that these GC-like CWs are not dynamically featureless; rather, they are weakly rotating stellar systems embedded in a still-coherent gaseous structure. 

\subsection{Globular cluster-like kinematics and densities}

We can investigate this idea further in the stellar $v_{\rm rot}$--$v/\sigma$ plane (top panel of Figure~\ref{fig:Rot3}), where the kinematic properties of the clusters are compared directly to those of local GCs.\footnote{We exclude the main bulge GC population from this comparison, since their present-day kinematics is likely to have been strongly shaped by the deep inner Galactic potential, repeated tidal shocks, and bulge/disc passages.} The observed GC population follows a clear sequence of increasing $v/\sigma$ with $v_{\rm rot}$ \citep[][]{Sollima:2019aa, Vasiliev:2021aa}, confined to low rotational velocities ($v_{\rm rot} \lesssim 10\,{\rm km\,s^{-1}}$). The CWs span a wide range in both quantities and overlap with this sequence at low $v_{\rm rot}$, showing that GC-like dynamical states can already be present at formation. At the same time, they extend to higher $v_{\rm rot}$ and $v/\sigma$, overlapping with part of the disc-cluster population.

The proto-GC interpretation for CGM-forming clusters is further supported by the green triangles, which show the proto-GC candidates from \citet{vandonkelaar:2023aa}. These were identified in the high-resolution $N$-body hydrodynamical zoom-in simulation \textsc{GigaEris} \citep[][]{Tamfal:2022aa} as gravitationally bound, baryon-dominated stellar systems forming in the CGM, predominantly along accreting gas filaments at $z \sim 4.4$. These systems lie close to the observed GCs. 

Taken together, these results show that clusters forming in the CGM can naturally reach GC-like dynamical states, although some CWs likely still need to lose angular momentum. This is not necessarily problematic, since both observations and theoretical studies suggest that stellar clusters can gradually lose rotational support through two-body relaxation, tidal interactions, mass loss, and dynamical heating over cosmic time \citep[e.g.][]{Ernst:2007aa, Bianchini:2018aa}. The systems presented here should therefore be interpreted as very early dynamical states rather than direct analogues of present-day GCs.

The bottom panels of Figure~\ref{fig:Rot3} connect these kinematic properties to the stellar surface density, $\Sigma_{\star}$. The bottom left-hand panel shows $\Sigma_{\star}$ as a function of $v_{\rm rot}$, where a subset of CWs overlaps again with the GC region, while disc clusters remain offset to higher $v_{\rm rot}$ and $\Sigma_{\star}$. The bottom right-hand panel shows $\Sigma_{\star}$ as a function of $v/\sigma$, where a small subset of CWs overlaps with the observed GC parameter space, but the population also extends to higher $v/\sigma$ and higher densities. Disc clusters, in contrast, overlap substantially with the CWs in $v/\sigma$, but remain offset from the GC population towards higher stellar surface densities.

Taken together, these panels reinforce the notion that CWs occupy a broad dynamical space, including both low-rotation systems with GC-like stellar surface densities and denser, more rapidly rotating systems that may instead evolve through runaway collapse towards IMBH seed formation \citep[][]{Gurkan:2004aa}. This interpretation is consistent with the picture developed in \citet{Donkelaar:2026aa}, where the densest CWs are expected to undergo runaway stellar collisions and form BH seeds. In this context, the low-rotation CWs discussed here occupy the opposite end of the population, combining lower densities and weaker rotational support, making them less favourable sites for runaway collapse and more natural proto-GC candidates. This suggests that the fate of CWs is set by both their internal density and angular momentum \citep[e.g.][]{Kruijssen:2015aa, Renaud:2017aa}, with the CGM environment providing the initial conditions for these different pathways.

\begin{figure}
    \centering
    \includegraphics[ trim={0cm 0cm 0cm 0cm}, clip, width=0.485\textwidth, keepaspectratio]{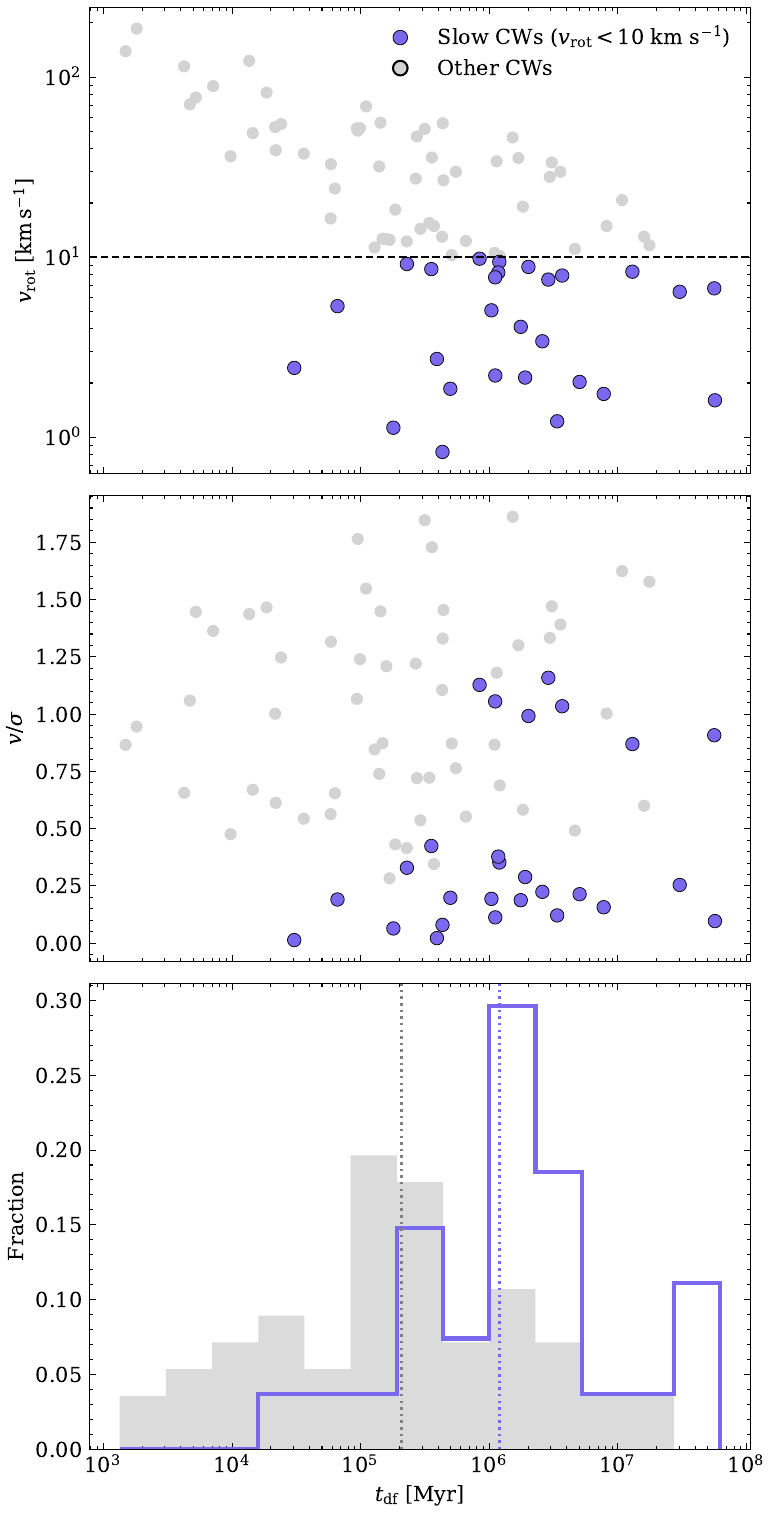}
   \caption{DF time-scale, $\tau_{\rm DF}$, for CWs split into slow rotators ($v_{\rm rot}<10\,{\rm km\,s^{-1}}$) and the remaining CW population. The top panel shows $\tau_{\rm DF}$ against rotational velocity, the middle panel shows $\tau_{\rm DF}$ against $v/\sigma$, and the bottom panel shows the normalized distribution. The dotted vertical lines in the bottom panel show the median values of each sample.}
    \label{fig:tdf}
\end{figure}

\subsection{Survival of the low-rotation cosmic wallflowers}

To estimate whether the CWs are expected to rapidly migrate towards the centre of the main halo, we compute the dynamical-friction (DF) time-scale following the same approach as in \citet{Donkelaar:2026aa}. We use the combined gas, stellar, and dark matter density profile of the main halo to calculate the enclosed mass, $M(<r)$, and the corresponding circular velocity,

\begin{equation}
    V_{\rm c} = \sqrt{\frac{G M(<r)}{r}} \,,
\end{equation}

\noindent where $r$ is the three-dimensional distance from the centre of Halo~0 and $G$ is the gravitational constant. The DF time-scale is then estimated as \citep[][]{Binney:2008aa}

\begin{equation}
    \tau_{\rm DF} =
    1.17\,\frac{V_{\rm c} r^2}{G M_{\rm cl}\ln\Lambda}\,,
\end{equation}

\noindent where $M_{\rm cl}$ is the baryon cluster mass within the stellar half-mass radius, and $\ln\Lambda$ is the Coulomb logarithm. This gives an approximate orbital-decay time-scale for clusters on roughly circular orbits in the background potential of the main halo. As in \citet{Donkelaar:2026aa}, these values should be interpreted as relative survival indicators rather than exact infall times, since radial motions, interactions with other substructures, and the time-dependent merger environment can shorten or lengthen the true migration time.

The low-$v_{\rm rot}$ CWs are the closest analogues of present-day halo GCs in the kinematic comparison above, but their long-term relevance depends on whether they can avoid rapid inspiral into the centre of the main galaxy. We therefore estimate the DF time-scale, $\tau_{\rm DF}$, for the CW population, separating slow rotators ($v_{\rm rot}<10\,{\rm km\,s^{-1}}$) from the rest of the sample. Figure~\ref{fig:tdf} shows that he absolute DF time-scales are long for both CW subsets, with most systems exceeding $\tau_{\rm DF}=10^4$~Myr. The slow rotators are shifted towards somewhat longer values, but this difference is small compared with the absolute time-scales. The slow rotators have a median galactocentric radius of $9.21$ kpc, compared with $6.27$ kpc for the remaining CWs. They also occupy the lower-mass end of the CW population, with median masses approximately two to four times lower than the rest of the sample. Since $\tau_{\rm DF}\propto r^2/M_{\rm cl}$, both their lower masses and somewhat larger galactocentric radii contribute to the modest shift towards longer time-scales.

The main result is therefore that rapid inspiral is unlikely for most CWs. Although the slow rotators are shifted towards somewhat longer $\tau_{\rm DF}$ than the remaining CWs, the difference between the two populations is modest. Together with their lower densities, this further suggests that the slow-rotating CWs are unlikely to follow the BH-seed formation pathway proposed for the densest CWs.

\begin{figure}
    \centering
    \includegraphics[ trim={0cm 0cm 0cm 0cm}, clip, width=0.485\textwidth, keepaspectratio]{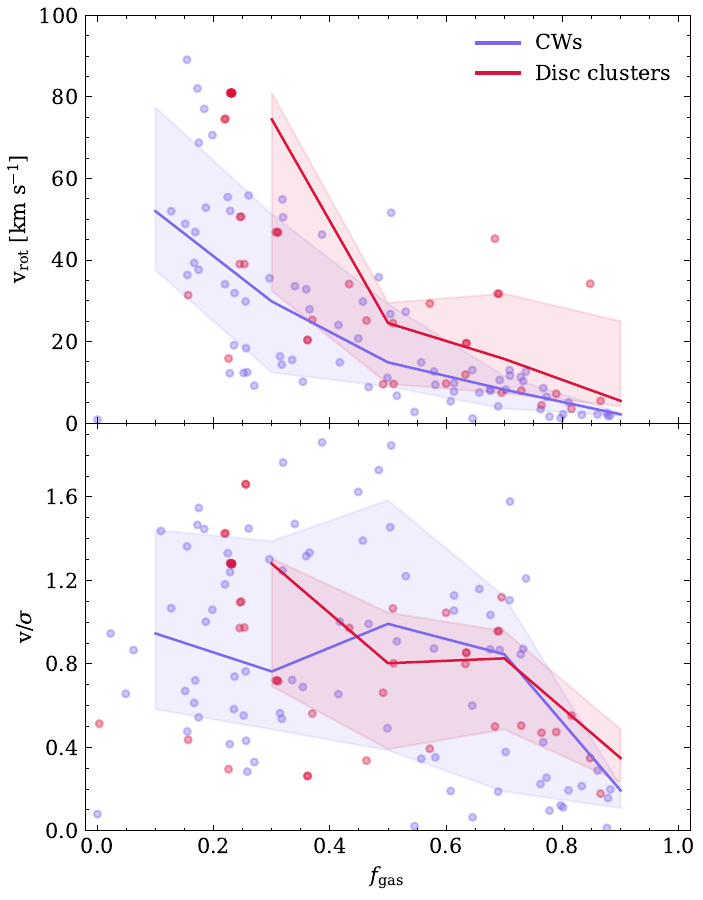}
    \caption{Gas fraction versus stellar kinematics for CWs (blue) and disc clusters (red) at the final snapshot. The top panel shows the stellar rotation velocity, $v_{\rm rot}$, and the bottom panel the rotational support, $v/\sigma$, as a function of gas fraction, $f_{\rm gas}$. Points indicate individual clusters, whereas the solid lines show the binned median in $f_{\rm gas}$, with shaded regions marking the 16th--84th percentile range, and only showing for bins containing at least five objects.}
    \label{fig:fgas}
\end{figure}

\subsection{Gas fraction}

Lastly, Figure~\ref{fig:fgas} shows that gas fraction is related to the stellar kinematics of both populations. For both CWs and disc clusters, increasing $f_{\rm gas}$ is associated with decreasing $v_{\rm rot}$, although the disc clusters remain shifted towards somewhat higher rotational velocities over much of the overlapping range. A similar tendency is visible in $v/\sigma$, particularly towards the most gas-rich systems.

For the CWs, the clearest trend is with $v_{\rm rot}$. Gas-rich CWs preferentially occupy the low-$v_{\rm rot}$ regime associated with the GC-like subset, while gas-poor systems extend to higher $v_{\rm rot}$ and more moderate $v/\sigma$. Gas fraction therefore does not define a simple evolutionary sequence, but traces part of the dynamical diversity within the CW population. If these CWs retain their gas and continue forming stars for longer, it would also naturally connect to the multiple-population and metallicity spread picture for GCs.

Nevertheless, the gas component may still be dynamically important at early times. For example, \citet{bobrick:2025aa} show that compact clusters can retain gas in their potential wells, with their subsequent evolution depending on the depth of the cluster potential. Although the gas in our CWs may have a different origin, this illustrates that a bound gas reservoir can remain dynamically relevant during early cluster evolution.

\section{Discussion and conclusions}\label{sec:disc}

The results presented in this work show that the dynamical state of star-forming clusters at $z\sim7.6$ is strongly shaped by their formation environment. By comparing clusters forming in galactic discs and in the CGM within \textsc{MassiveBlackPS}, we find the clearest difference in their rotational velocities: disc clusters are shifted towards higher $v_{\rm rot}$, while CWs preferentially occupy the low-rotation regime. Their $v/\sigma$ distributions overlap substantially, although the disc clusters are slightly shifted towards higher rotational support. This broad dynamical range provides a natural route for some CGM-forming clusters to reach GC-like states already at formation.

One possible explanation for this difference is that the two populations inherit different amounts of angular momentum from their parent gas flows. Disc clusters form within rotationally supported galactic discs and are therefore expected to originate from gas with substantial angular momentum. In contrast, CWs may preferentially form in lower-vorticity regions of the CGM, naturally producing lower stellar rotational velocities. While we do not directly measure the local gas vorticity in this work, the systematically lower $v_{\rm rot}$ of the CW population and their coherent gas-star kinematics are qualitatively consistent with this picture \citep[see also][]{Renzini:2025aa}. A detailed analysis of the local velocity field surrounding forming clusters will be explored in future work.

Crucially, a subset of the CWs occupies the same region of parameter space as observed GCs, combining low rotational velocities with stellar surface densities comparable to those of local systems. These clusters are not simply the low-mass tail of the overall population, but represent systems with distinct dynamical properties, indicating that GC-like conditions are already established at formation. This interpretation is further supported by the proto-GC candidates identified by \citet{vandonkelaar:2023aa}, which lie close to the GC locus while also forming in the CGM within the \textsc{GigaEris} simulation, demonstrating that such environments can directly produce GC-like systems. In contrast, clusters forming in discs at these high redshifts remain offset to higher $v_{\rm rot}$ and do not overlap with the local GC population, suggesting that they likely follow a fundamentally different evolutionary pathway.

The measured rotation of the CWs appears to trace coherent internal motion rather than numerical noise. Their gas and stellar rotational velocities closely follow each other, while their high co-rotating mass fractions point to a coherent, vorticity-like angular-momentum structure. For the low-$v_{\rm rot}$ CWs, this is particularly important: they rotate only weakly in stars, but remain embedded in coherent gas and have DF time-scales that are generally too long for rapid inspiral. These systems also tend to occupy the lower-mass end of the CW population and are found at somewhat larger galactocentric radii, which may contribute to the modest shift towards longer DF time-scales. This makes them unlikely to sink rapidly into the central galaxy, supporting their interpretation as plausible proto-GC candidates.

The CW population itself is also not homogeneous. Instead, it separates into two physically distinct regimes, as discussed in \citet{Donkelaar:2026aa}. One subset consists of lower-density, weakly rotating systems that overlap with the GC locus and represent the most plausible proto-GC candidates. The second subset is comprised of denser and more rotationally supported systems, occupying a region closer to the disc clusters. These systems are unlikely to evolve into present-day GCs and are instead natural candidates for runaway stellar collisions and BH seed formation  \citep[][]{Mayer:2025aa, Donkelaar:2026aa}. The generally long DF time-scales of the low-rotation systems further indicate that rapid inspiral is unlikely, although the difference relative to the remaining CW population is modest.

The connection between gas content and kinematics further reinforces this picture. Both populations show decreasing $v_{\rm rot}$ towards larger gas fractions, indicating that gas content is connected to the internal dynamical state in both formation environments. Gas-rich CWs are preferentially found in the weakly rotating regime and overlap most closely with the GC population, while gas-poor systems extend towards higher $v_{\rm rot}$ and greater rotational support. Gas fraction is therefore linked to the dynamical state of both populations, whereas formation environment produces an additional difference in their characteristic rotational velocities. Within the CW population, it also traces the different internal conditions of the low- and high-rotation systems. This behaviour is broadly consistent with recent models in which the retention and processing of gas within clusters plays a central role in their early evolution and the emergence of multiple stellar populations \citep[e.g.][]{bobrick:2025aa}.

Nevertheless, these results should be interpreted with some caution. Even at $\sim$pc-scale resolution, the CWs represent compact, star-forming systems at formation rather than fully evolved bound clusters. Their internal structure and kinematics are not fully converged, and the simulation does not follow the long-term collisional evolution required to assess their survival to $z=0$. In addition, it has been shown that the amount of small-scale fragmentation depends on the resolution of the cooling scale \citep[e.g.][]{Inoue:2015aa}, which may therefore also affect the detailed properties of the densest systems in this work.

Despite these limitations, our results provide a coherent framework linking cluster formation environment, gas content, internal dynamics, and survival time-scales to the origin of GCs. In this picture, only a subset of high-redshift clusters are likely to evolve into present-day GC analogues. These systems preferentially form outside galactic discs and are already relatively weakly rotating and gas-rich. Their long DF time-scales make rapid inspiral unlikely, while further
angular-momentum loss may bring their kinematics closer to those of present-day GCs. The majority of clusters, including those forming in discs, instead follow alternative evolutionary pathways. The emergence of GC-like systems is therefore not a generic outcome of clustered star formation, but the result of specific conditions that are already established at early times.

\section*{Acknowledgements}
We thank the reviewer, Oleg Gnedin, for the useful comments
that greatly improved this work. We also thank Alvio Renzini for fruitful discussions that helped inspire this work. FvD acknowledges support from the Herchel Smith Fellowship at the University of Cambridge. PRC acknowledges support from the Swiss National Science Foundation under the Sinergia Grant CRSII5\_213497 (GW-Learn) and from the Swiss State Secretariat for Education, Research and Innovation (SERI) through the SKACH grant. The simulations were performed on the Piz Daint and Alps/Eiger Cray supercomputers at the Swiss National Supercomputing Center (CSCS) under the uzh3 rolling allocation.

\section*{Data Availability}
The data underlying this article will be shared on reasonable request to the corresponding author.



\bibliographystyle{mnras}
\bibliography{example} 

\newpage

\appendix

\section{High-$v/\sigma$ cosmic wallflowers}\label{sec:CF}

We further examine the nature of the high-$v/\sigma$ CWs to assess whether they are consistent with a cluster population, and might not be any other astronomical object. In \citet{Donkelaar:2026aa}, an extensive cluster selection procedure was applied, combining \textsc{Amiga Halo Finder} \citep[][]{Gill:2004aa,Knollmann:2009aa} identification with strict cuts on size, mass, and baryon fraction, as well as visual inspection, to exclude dwarf galaxies and ensure that only compact, bound stellar systems are selected. All systems considered here satisfy these criteria and remain strongly baryon-dominated, with no evidence for associated dark matter subhaloes.

\begin{figure}
    \centering
    \includegraphics[ trim={0cm 0cm 0cm 0cm}, clip, width=0.48\textwidth, keepaspectratio]{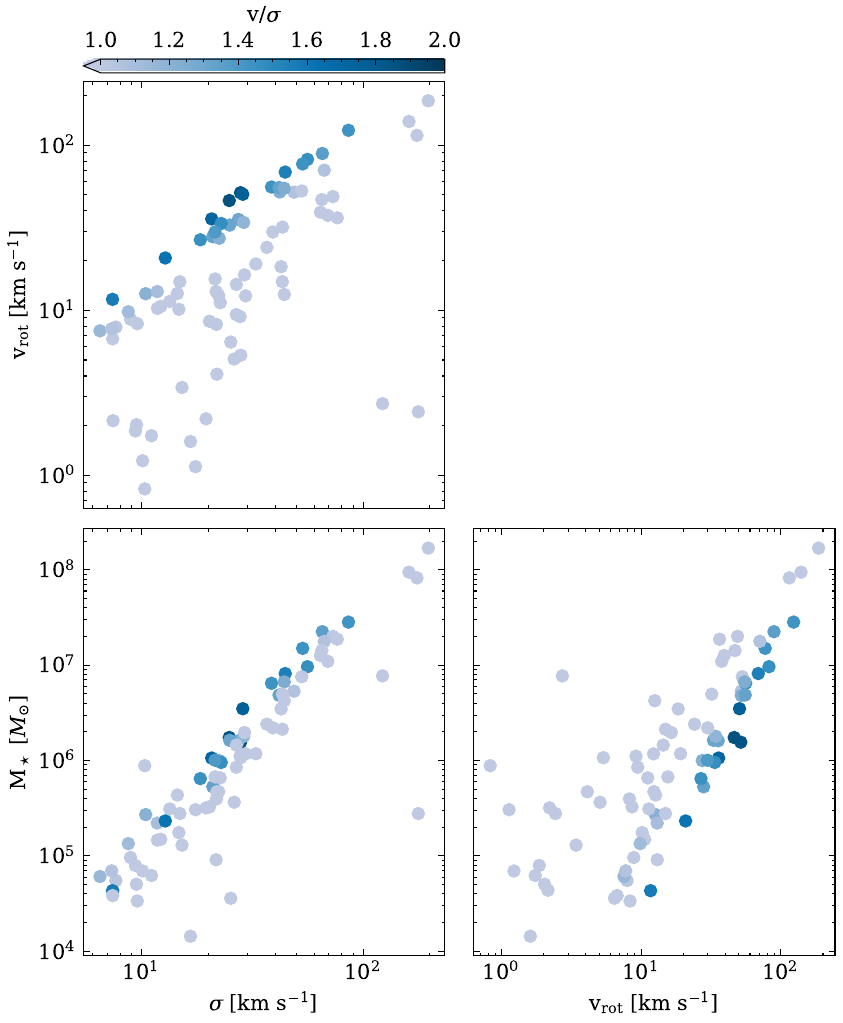}
    \caption{Rotational velocity, $v_{\rm rot}$ (top left), and stellar mass, $M_\star$ (bottom left), as a function of $\sigma_{\rm rot}$, and $M_\star$ as a function of $v_{\rm rot}$ (bottom right) for the CW population. Points are coloured by $v/\sigma$. The highest $v/\sigma$ systems do not exhibit distinct behaviour in either mass or rotational velocity, indicating that they do not represent a separate population.}
    \label{fig:mstar}
\end{figure}

Figure~\ref{fig:mstar} shows that the high-$v/\sigma$ systems do not stand out in terms of stellar mass, rotational velocity, or velocity dispersion. No clear correlation is found between mass and either $v_{\rm rot}$, $\sigma$, or $v/\sigma$, and the highest $v/\sigma$ systems do not occupy a distinct region of parameter space. This indicates that the high-$v/\sigma$ systems are not associated with dark matter subhaloes or mini-galaxies, but are consistent with being genuine cluster systems within the dynamical range of the CW population. 

\bsp	
\label{lastpage}
\end{document}